\documentclass[conference]{IEEEtran}

\usepackage{multirow}
\usepackage{graphicx}
\usepackage[table,xcdraw]{xcolor}

\begin{document}
\bstctlcite{MyBSTcontrol}

\title{Compressing and Comparing the Generative Spaces of Procedural Content Generators
}

\author{\IEEEauthorblockN{1\textsuperscript{st} Oliver Withington}
\IEEEauthorblockA{\textit{Queen Mary University of London} \\
London, UK \\
o.withington@qmul.ac.uk}
\and
\IEEEauthorblockN{2\textsuperscript{nd} Laurissa Tokarchuk}
\IEEEauthorblockA{\textit{Queen Mary University of London} \\
London, UK \\
laurissa.tokarchuk@qmul.ac.uk}
}

\maketitle
\makeatletter
\def\ps@IEEEtitlepagestyle{%
  \def\@oddfoot{\mycopyrightnotice}%
  \def\@oddhead{\hbox{}\@IEEEheaderstyle\leftmark\hfil\thepage}\relax
  \def\@evenhead{\@IEEEheaderstyle\thepage\hfil\leftmark\hbox{}}\relax
  \def\@evenfoot{}%
}
\def\mycopyrightnotice{%
  \begin{minipage}{\textwidth}
  \centering \scriptsize
  Copyright~\copyright~2022 IEEE. Personal use of this material is permitted. Permission from IEEE must be obtained for all other uses, in any current or future media, including\\reprinting/republishing this material for advertising or promotional purposes, creating new collective works, for resale or redistribution to servers or lists, or reuse of any copyrighted component of this work in other works by sending a request to pubs-permissions@ieee.org.
  \end{minipage}
}
\makeatother

\IEEEpubidadjcol
\begin{abstract}
The past decade has seen a rapid increase in the level of research interest in procedural content generation (PCG) for digital games, and there are now numerous research avenues focused on new approaches for driving and applying PCG systems. An area in which progress has been comparatively slow is the development of generalisable approaches for comparing alternative PCG systems, especially in terms of their generative spaces.  It is to this area that this paper aims to make a contribution, by exploring the utility of data compression algorithms in compressing the generative spaces of PCG systems. We hope that this approach could be the basis for developing useful qualitative tools for comparing PCG systems to help designers better understand and optimize their generators. In this work we assess the efficacy of a selection of algorithms across sets of levels for 2D tile-based games by investigating how much their respective generative space compressions correlate with level behavioral characteristics.  We conclude that the approach looks to be a promising one despite some inconsistency in efficacy in alternative domains, and that of the algorithms tested Multiple Correspondence Analysis appears to perform the most effectively.

\end{abstract}
\IEEEpeerreviewmaketitle

\section{Introduction}

Procedural Content Generation (PCG) for games, the algorithmic generation of digital artifacts, has in the past decade developed into a lively and diverse research field, with a proliferation of new works exploring novel implementations and use-cases of PCG systems. As the volume of new work has increased, it has become increasingly important to develop methods for comparing alternative generators and their outputs. Without them it is hard to identify when a new approach is a useful advance for the field, or for game designers to select the best approach for their purposes. For some forms of analysis this is more straightforward. PCG systems can be deployed in settings with quantifiable goals such as providing training material for Reinforcement Learning (RL) agents \cite{justesenIlluminatingGeneralizationDeep2018} \cite{risi2020}, or more niche ones like generating high-density weather resistant cities \cite{galanos2021}.  For use-cases of PCG where the purpose of the generated artifacts is to be consumed by players or designers then comparing PCG systems is much more complex, as the role the artifacts are playing is fundamentally a subjective and artistic one. This makes it hard to assess the extent to which a generator is achieving its intended purpose.

A central concept for when the goal is to compare PCG systems in terms of their output, is that of 'generative space'. This refers to the conceptual volume that represents ‘the theoretical space of all possible output of a generator \cite{guzdial2020}. When the goal is to compare PCG systems in terms of their output the underlying goal is to be able to draw useful comparisons between their generative spaces, but this can be extremely challenging. The total size of the spaces in terms of the number of possible artifacts can be extremely large, such as the over 18 quintillion procedurally generated planets that No Man Sky boasted at launch \cite{khatchadourian2015}. The relationship between the contents of a generative space and the parameterisation of the generator can also be unpredictable, especially if there are stochastic elements to the generation process. Furthermore, the applicability of a given generative space to a given domain can also be a subjective and artistic one. In some settings it might be acceptable to have undesirable or even non-functional artifacts in their generative space so long as the average output is diverse and interesting enough, such as when a filtering process is involved. For another, any unusable artifacts being present may be unacceptable, for example in settings where all generated content is 'necessary' \cite{togelius2011}. 

A common approach for aiding designers in understanding generative spaces is to convert the extremely high dimensional uninterpretable volume which contains the direct encodings of the generated artifacts into something lower dimensional and human understandable. The most common way of doing this in prior research on game level generation is Expressive Range Analysis (ERA) \cite{smith2010}. This approach visualises generative spaces by calculating and mapping emergent Behavioral Characteristics (BCs) of the generated levels such as heuristics for difficulty or aesthetic qualities. ERA has been widely adopted as a metric for qualitatively comparing different generative spaces (See  \cite{jadhav2021, zakaria2021, sarkar2021} for recent examples). However, it is also possible to directly compress the encoded representations themselves to produce similar low dimensional representations without the need to decide on BCs of interest as in ERA. This is commonly done as an intermediary step in a larger generative process, often to produce a low dimensional form of levels that can be understood by a neural network \cite{thakkar2019, sarkar2021, sarkar2020}. However we argue this direct compression can be more intrinsically valuable for producing useful representations of generative space which can help bridge the gap between designers and understanding of their generators.

The approach we explore in this paper is to compress high dimensional encoded representations of levels from a generative space to produce two dimensional projections that capture as much of the variance in the level set as possible. This projection which represents the underlying generative space can then be qualitatively understood and compared to alternatives in terms of the types and variety of levels that can be produced. If effective, this could let designers more easily understand and compare generative spaces without needing to make any decisions about the types of diversity which are of interest. We explore this approach in the context of generators for 2D tile-based games as there is a large amount of prior work and research that we can leverage such as level generators, pre-generated level corpuses \cite{boxobanlevels, summerville2016a} and research frameworks \cite{karakovskiy2012, perez-liebana2016}. By taking samples of encoded generated levels from a system or selection of systems and treating them as sets of variables in which each variable represents a portion of the level, we can then use dimensionality reduction algorithms to represent that set using a smaller number of new compressed variables. The goal is that this compression approach will make it significantly easier to understand and compare the types and variety of content that can be produced by alternative PCG systems in a way that is easy to configure and domain agnostic.

To experimentally explore this approach we assess four commonly used dimensionality reduction algorithms (PCA, SVD, MCA and T-SNE) in the domain of compressing the generative spaces PCG systems for three 2D tile-based games: Super Mario, Lode Runner and Boxoban \cite{boxobanlevels}, an open source version of Sokoban. To aid in the comparison between the alternative algorithms, we assess the extent to which diversity in the compressed low dimensional space correlates with diversity in terms of behavioral characteristics (BCs) of the game levels. We conduct analysis on the linear correlation between the distances between levels in the compressed generative space against the difference between their BCs. The more that there is a correlation between the two, the more credibly we can claim that we are compressing the generative spaces of PCG systems while preserving behavioral information which would be useful to game designers. While the idea of applying dimensionality reduction algorithms to encoded game levels is not itself novel,  we believe this is both the first to use it to compare alternative generative spaces, as well as the first to explore the correlation between the compressed space and the behavioral features of the levels.

The rest of this paper is laid out as follows. In Section \ref{rel_work} we discuss the most relevant related work and how this project builds on its ideas. In Section \ref{approach} we introduce and discuss the approach used, and the system we have implemented to assess it. In Section \ref{exp_design} we explain the experimental design of the experiments presented, and in Section \ref{results} we present the results from these experiments. In Section \ref{discussion} we discuss the implications of the results, as well as the limitations of the underlying approach and the future work that should be done to further explore it. In Section \ref{conclusion} we conclude that this appears to a promising approach for understanding and comparing generators which is worthy of further examination in alternative domains and configurations.

\section{Related Work}\label{rel_work}
The concept of generative space appears in a majority of works focused on PCG systems, either directly or indirectly. Depending on the researcher and the context, many different terms can be used to refer to a PCG system's generative space. They can be referred to as 'search spaces' in the context of generate-and-test PCG systems \cite{liapis2015, gravina2019}, or as 'possibility spaces' in work using stochastic PCG systems \cite{galanos2021, mason2019}. Researchers investigating Quality-Diversity (QD) algorithm based approaches for PCG often refer to 'behavioral space' as QD approaches rely on characterising generative spaces in terms of emergent artifact behaviors. In each case the underlying concept is largely the same. They are all different ways of conceptualising the total set of possible outputs from a PCG system. In this work we use the term 'generative space' as it is widely understood as well as generalisable to different domains.

The other concept used frequently in this paper is 'Behavioral Characteristic' (BC). This term is commonly used in PCG works based on Quality-Diversity search (See  \cite{sarkar2021, charity2020, fontaineCovarianceMatrixAdaptation2020} for recent examples) and it refers to emergent characteristics of generated artifacts which can be quantified to motivate the search for output diversity. Similar concepts often appear in PCG research under other names such as behaviors \cite{alvarez2019}, or more simply as 'metrics' \cite{cook2021, herve2021}. These BCs can be derived from the encoded artifacts directly \cite{smith2018, cook2021}, or derived from simulated play by an agent \cite{fontaineCovarianceMatrixAdaptation2020, khalifa2019}. 

The most prevalent method which aims to aid with the understanding and analysis of full generative spaces is Expressive Range Analysis (ERA). ERA was introduced by Smith and Whitehead in 2010 \cite{smith2010} and has since become a dominant method for understanding and comparing the generative spaces of PCG systems. To use it a designer selects two or more BCs of the generated levels which are then calculated for a sample of generated levels. They can then be visualised in lower dimensional space, typically on a 2D graph or heat map, allowing designers to visualise the generative spaces of their PCG systems in terms of BCs which are of most interest. This can have many benefits, such as highlighting where BCs are in conflict with each other and how different parameterisations of the same generator change the location and size of the resultant generative space in BC space \cite{smith2018, cook2021}. It is also used as a heuristic for comparing the output diversity of alternative generators \cite{jadhav2021}.

The primary limitations of ERA are that it can only be used to visualise two dimensions of diversity simultaneously, and that it can be challenging to determine and quantify what diversity is of interest. The first limitation can be offset using the approach of Summerville \cite{summerville2018} who used Corner Plots to visualise multiple BCs simultaneously, though the majority of works using ERA still opt to use sets of 2D visualisations based on two BCs for ease of readability. The second is more problematic as it speaks to the subjectivity of analysing the underlying artifacts. The recent work of Herve and Salge partly mitigates this weakness by exploring the relationship between commonly used BCs in PCG and expert evaluations of game content in the game Minecraft \cite{herve2021}. Their finding that there was significant correlation between perceptual differences between game artifacts and commonly used BCs bolsters the use of ERA, as well as the use of BCs as a heuristic for generative space diversity. However it does not address the issue that BCs and the heuristics that assess them need to be redesigned for each new game domain. As we will discuss later in this paper, the hope is that the approach presented here can realise many of the benefits of ERA while mitigating or avoiding these limitations. 

The approach discussed in this paper relies on applying dimensionality reduction algorithms to representations of game levels. This idea has been used as a preliminary or intermediary step in several pieces of PCG research which were direct inspirations for this work. In 'Sampling Hyrule' from Summerville and Mateas \cite{summerville2015} principal component analysis (PCA), a widely used dimensionality reduction algorithm, was used to compress representations of Zelda levels to construct a low dimensional representation of the space which could be sampled from to generate new levels. In 2018 Justesen et al used PCA to visualise how their generated level sets were distributed in relation to levels from the original games \cite{justesenIlluminatingGeneralizationDeep2018}. Variational Auto-Encoders, a neural network designed for dimensionality reduction,  have also become widely used in PCG for level generation \cite{snodgrassMultiDomainLevelGeneration2020, jadhav2021}. These works and our own relate back to the landmark paper 'Eigenfaces for Recognition' \cite{turkEigenfacesRecognition1991}, which found that PCA applied to raw image data of human faces could be used as the basis for accurate facial recognition software using only eight new variables. The insight that image data containing thousands of variables can be compressed while maintaining real world useful information adds weight to the idea that a similar approach could work for compressing generative spaces. 

This work is also closely related to the emerging and popular subfield of Machine Learning-based PCG approaches for game levels, commonly referred to as PCGML (See  \cite{liuDeepLearningProcedural2021} for a recent overview of this field). These approaches use neural networks to learn from sets of game levels and generate new ones. These techniques have been applied to many diverse goals, such as reproducing the style of expert designers \cite{volzEvolvingMarioLevels2018}, learning user preferences \cite{schrum2020} and generating new levels for unseen games \cite{jadhav2021}. These works are related to this one in two key ways. Firstly, they typically aim to extrapolate useful information about game levels directly from their representations, and their success adds credibility to the idea that encoded forms of game levels contain sufficient real-world useful information about their form and function. Secondly, they are very relevant to the concept of generative space. PCGML can be conceptualised as a process of learning to reproduce a generative space in the case of training directly from sets of game levels, or as the process of producing an ideal generative space from diverse inputs as in the case of learning from user preferences.

\section{Approach}\label{approach}

The goal for this approach is to represent sets of generated levels from a PCG system in a compressed two dimensional space, while maintaining enough information about the levels such that levels close together in the compressed space have similar BC values. To achieve this we apply dimensionality reduction algorithms to sets of game levels to create new uncorrelated variables composed out of combinations of the variables that compose the encoded level representations. We can then select the two new variables which explain the most variance in the underlying level data and reproject the level set in this space, giving us a two dimensional visualisation of the original high dimensional generative space.  This should help designers to answer questions such as:
\begin{itemize}
\item Whether a pair of generators produce similar levels
\item What kind of outliers are present in a generative space
\item What effect re-parameterisation of a generator is having on its output
\end{itemize}
The two requirements for this approach to be applied to set of levels are that every level be the same size and that they be assembled out of discrete parts in which each part can have one of a discrete set of values. Asides from these two requirements the approach is intended to be content agnostic and applicable to alternative content representations without significant configuration or domain specific tweaking.  

The high level steps of the system we use for applying and validating the approach are as follows:

To start, sets of levels are produced or sourced from each system that we want to evaluate which serve as representatives of the underlying generative space that they came from.  In this iteration of the system designed to work with tile-based 2D game levels, each level is loaded as a 2D matrix of characters in which each location in the matrix represents a tile in the level, and each character represents the corresponding tile type (i.e a solid block, empty space etc) that appears at that location.

The next step is to flatten the encoded levels into one dimensional arrays. If the compression algorithm being used uses categorical data then this is done in a single step, with each row of the character matrix combined horizontally into a single, ordered row. For algorithms which require numeric data we compress them into a 1D one-hot matrix in two steps. First, the character matrix is converted into a 3D one-hot matrix of size height x width x number of tile types, with every value set to 0 apart from those which indicate the tile type which appears at each location, which are set to 1. This 3D matrix can then be flattened to 1D in the same way as the categorical data. This one-hot conversion is the same that is used in many GAN-based PCGML works \cite{volzEvolvingMarioLevels2018, schrum2020}. The full set of 1D representations can then be stacked on top of each other to give a 2D matrix in which every row represents a level, and every column represents a location in the original level, or location and tile type in the case of one-hot encoding. 

The compressed 2D matrix representing all levels to analyse is now ready to have a dimensionality reduction algorithm applied. In this work we implement and compare four different algorithms: Single Value Decomposition (SVD), Principle Component Analysis (PCA), Multiple Correspondence Analysis (MCA) and T-distributed Stochastic Neighbor Embedding (T-SNE).  While they all operate differently (See Section \ref{comp_algos}), they are all designed to uncover underlying structures and dimensions in data so that it can be modelled using a new, smaller set of variables. The original data can then be reprojected using the top two most explanatory new variables produced by the respective algorithms. We note that all algorithms tested apart from T-SNE quantify the amount of variance explained by the generated variables. In typical uses of dimensionality reduction algorithms this is extremely important as it indicates how much of the mathematical variance of the underlying data set is being captured by the top n new variables. However, in this work we are interested in the behavioral difference between the generated levels, not in mathematical variance in their representations. As a result it is not the mathematical variance explained that we report on, but the amount that variance in the projected 2D space correlates with BC variance.

With the generative space visualisations now generated we can assess how effective each compressed projection is at capturing behavioral information about the levels. To assess this we calculate the linear correlation between BCs of the levels and the levels' relative positions in the compressed space using Spearman's rank coefficient. The claim we make is that the more that proximity between levels in the compressed space correlates with proximity in the levels's BC values the more credibly we can claim that the compression is conserving useful behavioral information about the levels and the more we are realising the benefits of ERA without many of its limitations. To calculate this we take every possible pair of levels and calculate the distance between their locations in the compressed spaces, and the difference between the values for commonly used BCs like number of enemies and linearity.  We then look for correlation between the two values by calculating Spearman's $\rho$, which we then use as our heuristic for the performance of the compression algorithm that produced the compressed space.

\section{Experiment Design}\label{exp_design}
In this section we provide an overview of the experimental design used in this work, as well as the justifications for the design decisions in the current implementation. The system is implemented in Python and is available on GitHub at https://github.com/KrellFace/Generative-Space-Compression

\subsection{Compression Algorithms}\label{comp_algos}

We implement four data compression algorithms in this work: PCA, SVD and MCA.  Kernal PCA \cite{romdhani1999} was also implemented but was found to not meaningfully outperform PCA in any configuration. For PCA we use the standard implementation provided by sklearn in its decomposition module. For SVD we use the TruncatedSVD implementation from the same module, though we note that this works identically to standard SVD. For T-SNE we use the implementation from the manifold module of sklearn and for MCA we use the implementation from Max Halfords prince module found at: (https://github.com/MaxHalford/prince). These four algorithms were chosen for several reasons, including their ease of implementation and the wide number of domains in which they have demonstrated usefulness \cite{ayesha2020}. More specifically, PCA was chosen as it has demonstrated utility at compressing game levels in prior works \cite{justesenIlluminatingGeneralizationDeep2018, summerville2015} and it is commonly accepted as 'the most important linear dimensionality reduction technique' \cite{vandermaaten2009}. SVD was chosen as it operates similarly to PCA except without first centering the data and we wanted to observe the influence of this on the final data. MCA is used as it is regarded as the categorical data counterpart to PCA \cite{greenacre1987}, and seeing as the levels are encoded as categorical rather than continuous data it could allow for similar analysis while avoiding a pre-processing step of converting to a one-hot matrix. Finally, T-SNE has also demonstrated utility in game level analysis \cite{jadhav2021} and is regarded as well suited specifically for visualising high dimensional datasets with nonlinear structures \cite{maaten2008} . However, there are many alternative algorithms with similar goals and outputs that could be used here \cite{ayesha2020}, and future work could usefully explore these further.

\subsection{Game Domains}

This work assesses generative spaces from three different game domains: Super Mario, Sokoban and Lode Runner, using open source level corpuses for each game.

For Super Mario we make use of the Mario AI Benchmark, an open source platform for AI research based on Super Mario \cite{karakovskiy2012}. The most up to date version of the platform can be found at (https://github.com/amidos2006/Mario-AI-Framework) courtesy of Ahmed Khalifa, and it comes packaged with pre-generated level sets from a selection of 9 generators which were generated as part of the work of Horn et al \cite{horn2014}, along with 15 levels from the original game. Each generated set comes with 1000 generated levels. All 9000 generated levels are used in our analysis, while the 15 from the original game are excluded as they are of varying sizes unlike the generated sets which are all 16 by 200 cell grids. We use a simplified encoding system in which every tile value is mapped to one of five types: Empty Space, Enemy, Solid, Pipe and Reward. A variety of generative approaches are represented in the level sets used \cite{horn2014}, and the Super Mario levels are the largest tested in terms of tile count by a significant margin. This should present the generative space compression approach with a challenge distinct from the other two domains, in which relatively distinct generative spaces are localised in large and sparse high dimensional spaces.

For Sokoban we use level sets provided as part of Guez et al's research into Boxoban using an open source variant called Bokoban \cite{boxobanlevels}. Three sets of levels are provided: 'unfiltered', 'medium' and 'hard'. The unfiltered set were generated by Guez et al using the approach of Racaniere et al \cite{racaniere2017}, and the other two were generated and selected the approach of Guez et al explained in  \cite{guez2019}. For the medium set 500,000 levels are available, stored in sets of one thousand in individual text files. For the unfiltered set one million levels are available, also stored in individual text files containing a thousand levels. The hard set contains 3,332 levels.  We use the same encoding used in the original work, with the only tile types available being solid block, empty space, pushable block, goal and player spawn location. The medium and hard sets were selected based on the failure of trained reinforcement learning agents at solving them. This means that the primary difference between the sets is not in the generative approach used in making them, but in the difficulty of solving them. This provides an interesting challenge for compressing the generative space of the three sets, as while they present significant variety in the gameplay experience provided in terms of their difficulty, they were all generated using a similar underlying approach before being filtered based on difficulty. This combined with the relatively small encoding sizes, with the smallest total size and tile variety of the three game domains, makes them an appealing challenge for testing this approach. 

For Lode Runner we assess only the 150 levels found in the original game. These are retrieved in an encoded form from the Video Game Level Corpus (VGLC) \cite{summerville2016a}, an open source repository of encoded tile-based game levels. We note that this means in this case we are not in fact assessing a generative space as this set was hand authored rather than coming from a generator. However, as there is little conceptual difference in compressing a generative with compressing a space of hand authored levels we feel that this will still be a valuable experiment. The small size and high aesthetic diversity of the level set also provide a different challenge to the compression technique than that provided by the other two game domains. We use the same encoding system as the VGLC in this work.

\subsection{Level Set Selection}

For the Super Mario and Boxoban domains, a subset of 4,000 levels are randomly selected for each experimental run and this selection is evenly distributed between the nine individual generator sets for Super Mario, and the three sets for Boxoban. A subset is taken for these domains to limit the computational resources required. A large sample size was chosen to increase the credibility of the Boxoban and Super Mario results as a larger set is presumed to be more representative of the underlying generative spaces than a small sample would be. However, initial experimentation as well as the Lode Runner results suggest that lower sample sizes could still produce effective results. For Lode Runner the full 150 levels are used in every run.

\subsection{Behavioral Characteristics}

For each game domain we calculate between two and three BCs (See Table \ref{table_bcs}). Each BC was selected for both being quick to calculate based on an encoded representation. Both linearity and enemy count have appeared frequently in prior work using ERA \cite{smith2010, shaker2012} and have also been found to reflect player perceptions \cite{summerville2017}. Contiguity, a measure which rewards solid blocks being adjacent, has has appeared in prior work using QD algorithms in PCG for tile-based games \cite{withington2020} and is intended to be a heuristic for how restricted player movement is in the level. Empty space amount, and other similar block count or ratio based BCs have also appeared in prior works implementing ERA \cite{jemmali2020}. 

\begin{table}[!t] 
\renewcommand{\arraystretch}{1.3} 
\caption{Behavioral Characteristics}
\label{table_bcs} 
\centering 
\begin{tabular}{|l|l|l|}
\hline
Game        & BC          & How Calculated (Count)              \\ \hline
Mario       & Empty Space (ES) & Empty Tiles                       \\
            & Linearity (Lin)   & Horizontally Adjacent Solid Tiles \\
            & Enemy Count (EC) & Enemy Tiles                       \\ \hline
Sokoban     & Empty Space (ES) & Empty Tiles                       \\
            & Contiguity (Contig)  & Adjacent Solid Tiles         \\ \hline
Lode Runner & Empty Space (ES) & Empty Tiles                       \\
            & Linearity (Lin)  & Horizontally Adjacent Solid Tiles  \\
            & Enemy Count (EC) & Enemy Tiles                       \\ \hline
\end{tabular}
\end{table}

\subsection{Linear Correlation}
For every combination of compression algorithm and game domain BC we calculate the difference between the values for every pair of levels in each set. For the compression algorithm space this is the vector distance between the level pairs respective position in the compressed projection. For the BCs we take the absolute difference between the values for the level pair. We then calculate the linear correlation between these values for every level pair, using Spearman's $\rho$ as we do not expect there to be a normal relationship between the two sets of values.

\subsection{Number of Runs}
To account for the influence of the random selection of levels for Super Mario and Boxoban, as well as the stochastic nature of T-SNE, ten runs are conducted for each game and algorithm.

\subsection{Computational Resources Used}
All experiments were run on an Intel i5-10310U CPU using a single thread and took approximately four hours to complete.

\section{Results}\label{results}


\begin{figure}[h]
\centering
\includegraphics[width=3.5in]{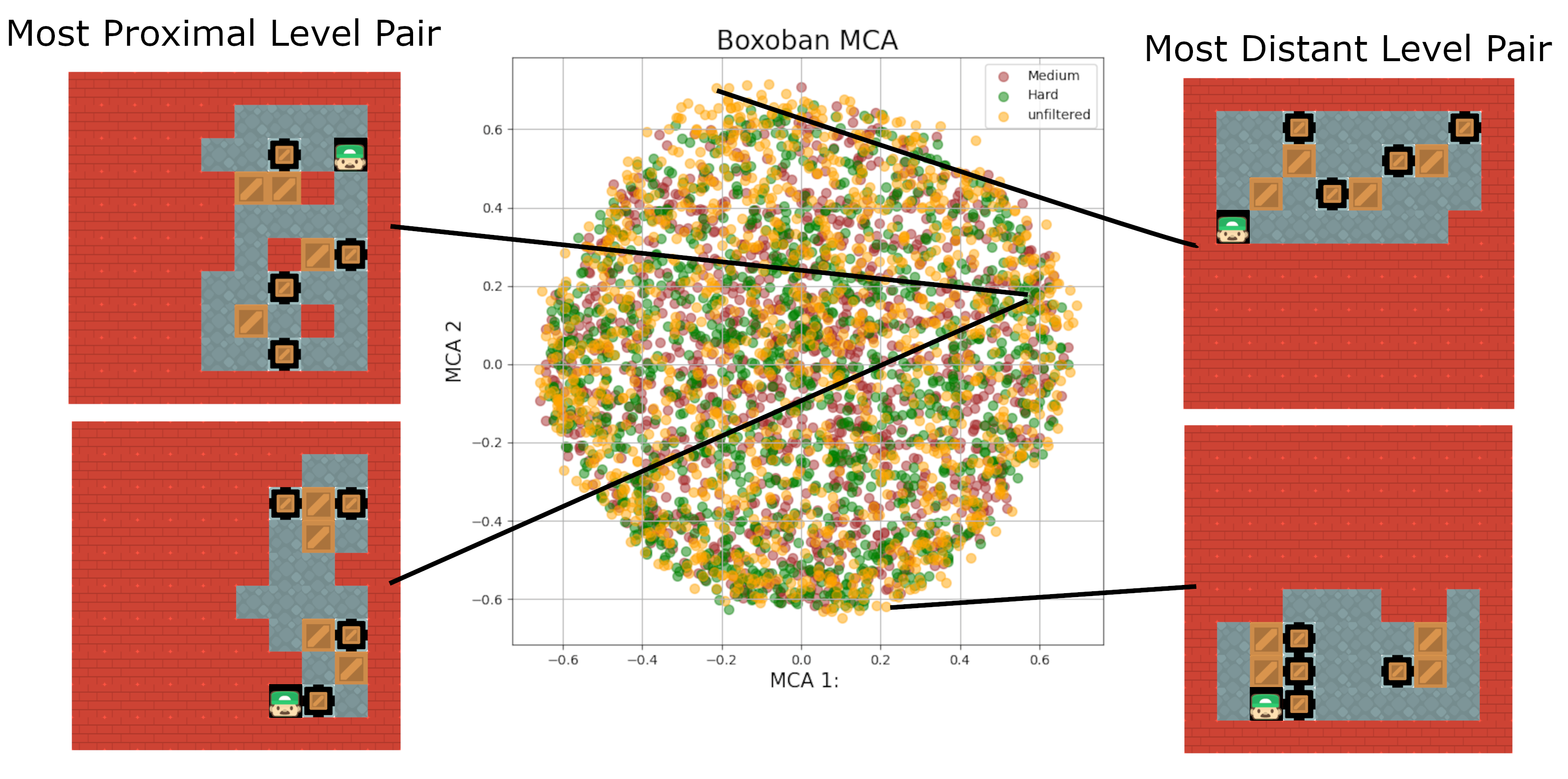}
\caption{Best Individual Boxoban Compression. T-SNE Run 9. Average BC Compression Correlation: 0.0380. Presented with the most proximal and most distant pair of levels in the compressed space}
\label{fig_bestboxoban}
\end{figure}

\begin{figure*}
\centering
\includegraphics[width=6in]{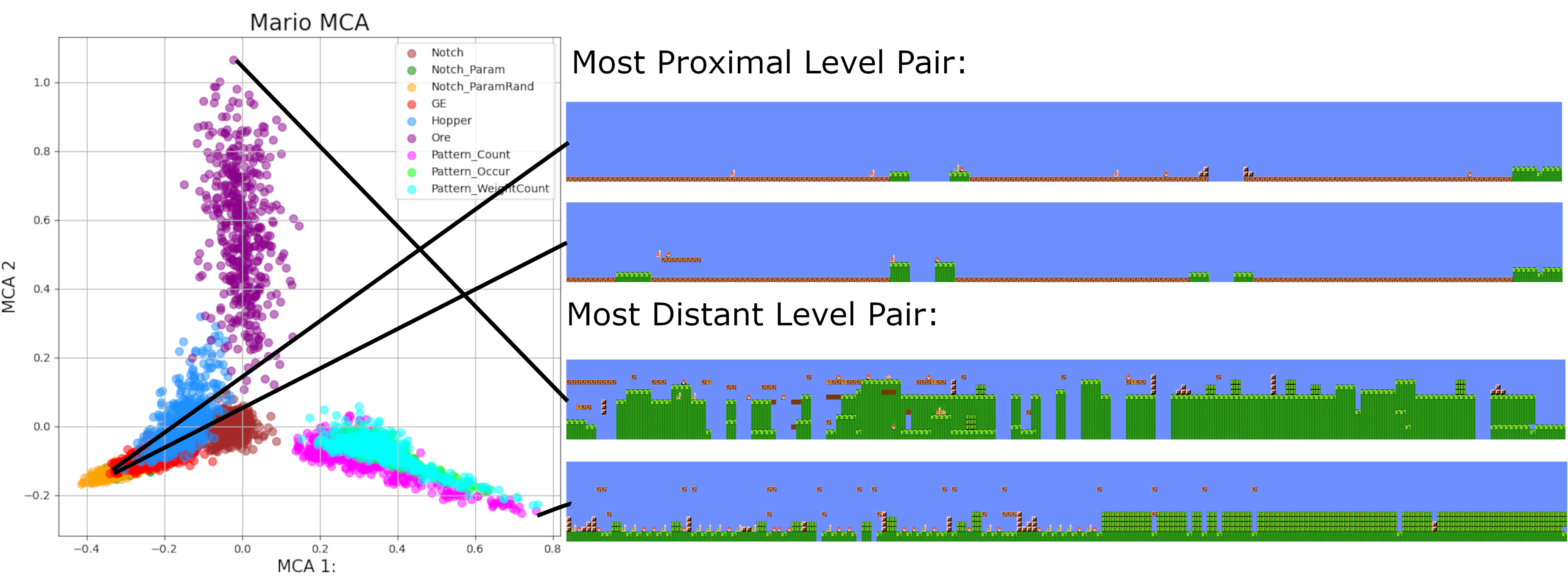}
\caption{Best Individual Super Mario Compression. MCA - Run 7. Average BC Compression Correlation: 0.497. Presented with the most proximal and most distant pair of levels in the compressed space}
\label{fig_bestmario}
\end{figure*}

\begin{figure}
\centering
\includegraphics[width=3.5in]{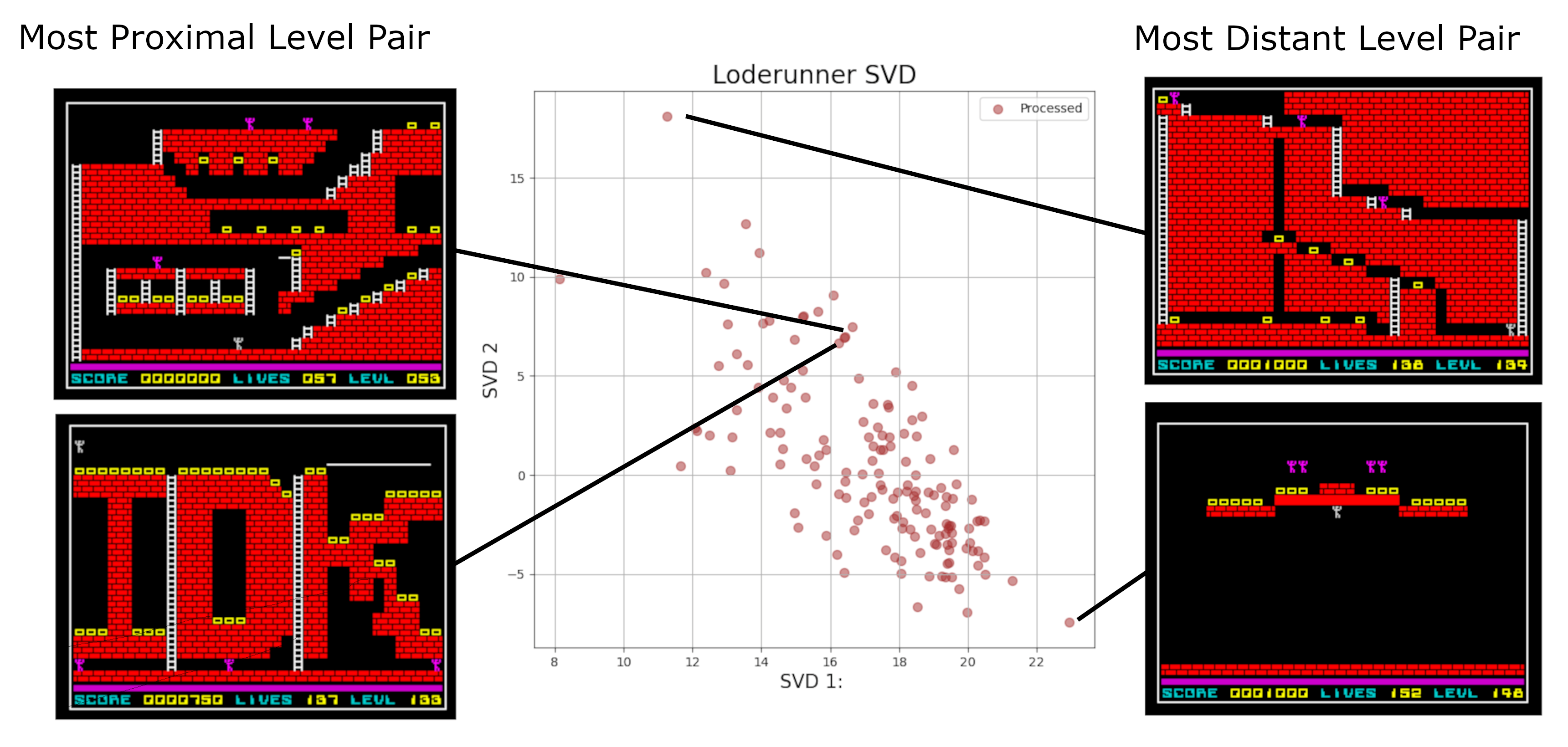}
\caption{Best Individual Lode Runner Compression. SVD Run 4. Average BC Compression Correlation: 0.461. Presented with the most proximal and most distant pair of levels in the compressed space}
\label{fig_bestlr}
\end{figure}

\begin{table*}
\caption{Average Spearman's Rhos and Associated P Values For Each Compression Algorithm and Game BC. Presented as Mean Avg +- StdDev. Best values for each Game BC Highlighted in Bold}
\begin{tabular}{|l|l|ll|ll|ll|ll|}
\hline
           &        & PCA                                                                &              & SVD                                                                &            & MCA                                                                &              & T-SNE                                                              &              \\ \cline{3-10} 
           &        & \multicolumn{1}{l|}{Spearman's $\rho$}                             & P Value      & \multicolumn{1}{l|}{Spearman's $\rho$}                             & P Value    & \multicolumn{1}{l|}{Spearman's $\rho$}                             & P Value      & \multicolumn{1}{l|}{Spearman's $\rho$}                             & P Value      \\ \hline
Mario      & ES     & \multicolumn{1}{l|}{\cellcolor[HTML]{32CB00}0.503$\pm$0.011}          & 0$\pm$0         & \multicolumn{1}{l|}{\cellcolor[HTML]{32CB00}0.530$\pm$0.005}          & 0$\pm$0       & \multicolumn{1}{l|}{\cellcolor[HTML]{009901}\textbf{0.765$\pm$0.003}} & 0$\pm$0         & \multicolumn{1}{l|}{\cellcolor[HTML]{34FF34}0.303$\pm$0.006}          & 0$\pm$0         \\ \cline{2-10} 
           & Lin    & \multicolumn{1}{l|}{\cellcolor[HTML]{34FF34}0.417$\pm$0.005}          & 0$\pm$0         & \multicolumn{1}{l|}{\cellcolor[HTML]{67FD9A}0.382$\pm$0.006}          & 0$\pm$0       & \multicolumn{1}{l|}{\cellcolor[HTML]{34FF34}0.497$\pm$0.006}          & 0$\pm$0         & \multicolumn{1}{l|}{\cellcolor[HTML]{34FF34}\textbf{0.494$\pm$0.005}} & 0$\pm$0         \\ \cline{2-10} 
           & EC     & \multicolumn{1}{l|}{\cellcolor[HTML]{34FF34}\textbf{0.330$\pm$0.003}} & 0$\pm$0         & \multicolumn{1}{l|}{\cellcolor[HTML]{34FF34}0.301$\pm$0.002}          & 0$\pm$0       & \multicolumn{1}{l|}{\cellcolor[HTML]{34FF34}0.295$\pm$0.004}          & 0$\pm$0         & \multicolumn{1}{l|}{\cellcolor[HTML]{34FF34}0.288$\pm$0.006}          & 0$\pm$0         \\ \hline
Boxoban    & ES     & \multicolumn{1}{l|}{\cellcolor[HTML]{9AFF99}0.024$\pm$0.010}          & 0$\pm$0         & \multicolumn{1}{l|}{\cellcolor[HTML]{9AFF99}0.034$\pm$0.011}          & 0$\pm$0       & \multicolumn{1}{l|}{\cellcolor[HTML]{9AFF99}\textbf{0.049$\pm$0.010}} & 0$\pm$0         & \multicolumn{1}{l|}{\cellcolor[HTML]{9AFF99}0.026$\pm$0.010}          & 0$\pm$0         \\ \cline{2-10} 
           & Contig & \multicolumn{1}{l|}{\cellcolor[HTML]{9AFF99}0.019$\pm$0.010}          & 0.017$\pm$0.052 & \multicolumn{1}{l|}{\cellcolor[HTML]{9AFF99}0.032$\pm$0.010}          & 0$\pm$0       & \multicolumn{1}{l|}{\cellcolor[HTML]{9AFF99}\textbf{0.046$\pm$0.010}} & 0$\pm$0         & \multicolumn{1}{l|}{\cellcolor[HTML]{9AFF99}0.020$\pm$0.009}          & 0.011$\pm$0.034 \\ \hline
Loderunner & ES     & \multicolumn{1}{l|}{\cellcolor[HTML]{32CB00}0.656$\pm$0}              & 0$\pm$0         & \multicolumn{1}{l|}{\cellcolor[HTML]{009901}\textbf{0.818$\pm$0.000}} & 0$\pm$0       & \multicolumn{1}{l|}{\cellcolor[HTML]{32CB00}0.582$\pm$0.000}          & 0$\pm$0         & \multicolumn{1}{l|}{\cellcolor[HTML]{9AFF99}0.133$\pm$0.025}          & 0$\pm$0         \\ \cline{2-10} 
           & Lin    & \multicolumn{1}{l|}{\cellcolor[HTML]{34FF34}0.440$\pm$0}              & 0$\pm$0         & \multicolumn{1}{l|}{\cellcolor[HTML]{32CB00}\textbf{0.557$\pm$0.000}} & 0$\pm$0       & \multicolumn{1}{l|}{\cellcolor[HTML]{34FF34}0.440$\pm$0.000}          & 0$\pm$0         & \multicolumn{1}{l|}{\cellcolor[HTML]{9AFF99}0.046$\pm$0.032}          & 0.036$\pm$0.115 \\ \cline{2-10} 
           & EC     & \multicolumn{1}{l|}{\cellcolor[HTML]{FFCCC9}-0.011$\pm$0}             & .266$\pm$0.001  & \multicolumn{1}{l|}{\cellcolor[HTML]{9AFF99}0.008$\pm$0.000}          & 0.408$\pm$0.0 & \multicolumn{1}{l|}{\cellcolor[HTML]{9AFF99}0.008$\pm$0.000}          & 0.390$\pm$0.005 & \multicolumn{1}{l|}{\cellcolor[HTML]{9AFF99}\textbf{0.042$\pm$0.042}} & 0.121$\pm$0.260 \\ \hline
\end{tabular}
\end{table*}

\section{Discussion}\label{discussion}

Overall we would describe the experimental results as promising but inconsistent. Across all game domains the highest performing compression algorithms produced compressed spaces which correlated with the behavioral characteristics assessed. However, there were areas in which the approach struggled to produce meaningful compressions. For the Enemy Count BC for Lode Runner the approach was unable to produce compressed spaces that correlated despite the approach otherwise performing well in that domain, and in the Boxoban domain the approach under-performed across all BCs. 

In terms of individual algorithm performance, MCA appeared to perform the most effectively out of the algorithms tested, producing the strongest correlations with BCs in 3 out of 8 domains. In this regard it only marginally outperformed SVD and T-SNE, which performed the best in 2 out of 8 domains. It also notably under-performed in specific domains, such as the Enemy Count BC for Mario, a phenomenon that requires more investigation to understand. In general while MCA performed the best, we would argue that the inconsistency of results should motivate more experimentation before alternatives are rejected.

Across the game domains assessed there was significant variance in performance of the approach. In both Super Mario and Lode Runner domains generative spaces which correlated substantially with BCs were found. In the Boxoban domain the performance for all algorithms and both BCs was significantly lower, with the best performing compression algorithm producing average Spearman's correlation coefficients of $<$0.05. 

We believe there are two possible reasons for the Boxoban under-performance. Firstly, the levels may be too similar to each other. Though 5000 levels were assessed for each run all three sets come from a similar generative approach and share features such as having exactly four movable boxes and goals. This is supported by the generative space visualisation (Fig. \ref{fig_bestboxoban}) which indicates substantial overlap between the three level sets, in contrast with the Super Mario visualisation (Fig. \ref{fig_bestmario}) in which the levels from different generators occupied different spaces. The primarily difference between the three sets being their difficulty may simply not be present or detectable using the approach presented. The second possible reason is that the approach may perform worse in domains with smaller level encodings. Boxoban had the lowest number of tile types at only 5, and the levels were the smallest tested at only a 10 by 10 grid. More experimentation is required to explore whether there's a correlation between size of encoded level and the efficacy of the generative space compression.

Furthermore, the Boxoban results highlight a general weakness with the approach, which is the inability to detect structural similarities which appear in different locations of the level. To a human observer, the pair of levels which were flagged as being the most dissimilar in the compressed space (Fig. \ref{fig_bestboxoban}) appear to be substantially similar. However, as the similar structures appear in two different halves of the level they are incorrectly identified in the compressed generative space as being completely dissimilar. 

As noted the approach also under-performed in the leniency BC within the Lode Runner game domain. We suspect this was a result of a combination of the relative paucity of input level data combined with the relative sparseness of enemies within the levels. The Lode Runner input set is both small at only 150 levels. Therefore it follows that the large scale structural differences between the levels would account for the majority of the variance within the set, leading to the low correlation with leniency due to the relatively low influence of enemy placement on the overall variance.

\subsection{Limitations and Future Work}

A primary limitation of this work is its focus on 2D tile-based games. While the focus on this domain made sense pragmatically due to the abundance of prior work we could use, tile-based games only make up a small portion of the contemporary games market. It is also limited by the small number of game domains assessed, and the fact that only the Mario level corpus contained sets from different generators. Future work could benefit from exploring this approach in domains with more complex content representations, with content produced from a wider range of similar and dissimilar generators. A first step could be applying it to the generative spaces of Minecraft map generators, as it is a popular 3D game whose maps are comprised of typed chunks.

A practical limitation of the approach presented is the requirement that every encoded level have the same encoded size. Future work could explore the approach used by Summerville and Mateas which used a graph cutting procedure to scale up the smaller Zelda levels to the size of the larger ones \cite{summerville2016a}, though this necessarily involves creating new information that is not present in the original representation. Future work could usefully explore these and alternative methods for applying our approach to game levels of varying size. 

As discussed when noting the aesthetic similarity of the two Boxoban levels flagged as being most dissimilar (Fig. \ref{fig_bestboxoban}), our approach can be weak in detecting structural similarities between levels. Future work could explore the use of alternative compression systems such as Convolutional Neural Networks that handle location independent structural similarity.

In this work we used correlations between the compressed spaces and BCs as the heuristic for whether the compressions were capturing useful information about the game levels. However, it would be valuable to explore this further with user studies to examine whether proximity between levels in compressed generative space correlates with player perceptions of the similarity of the levels. If they do correlate this would be robust evidence that the generative space compression retains useful information about the generative spaces.

In future work it will also be important to explore how best to make use of our generative space projections, ideally in cooperation with experienced level designers. While the original motivation for this project was to be able to assess the output diversity of a generative system, if the projections are sufficiently information rich and human understandable they could be more widely useful. For example, future work could explore their utility as the basis for mixed-initiative design tools, or automatic generator parameter tuners by allowing designers to target specific areas of the projected space.

\section{Conclusion}\label{conclusion}

In this paper we have presented an approach for compressing the generative spaces of PCG systems using dimensionality reduction algorithms. The approach appears to be a promising basis for developing generative space visualisation and comparison tools as despite its simplicity it was able to produce generative space projections which correlated significantly with behavioral features. Of the algorithms we tested MCA performed the most reliably, but not by a margin that eliminates alternatives from further investigation. Though more work is required to confirm the efficacy of this approach in more complex spaces, we hope this work could form the basis for new qualitative tools for aiding designer understanding of the generative spaces of PCG systems.

\section*{Acknowledgments}

This work was supported by the EPSRC Centre for Doctoral Training in Intelligent Games \& Games Intelligence (IGGI) [EP/S022325/1]. Thanks to Marko Tot and Michael Saiger for their helpful advice on an earlier draft of this work. 

\bibliographystyle{IEEEtran}
\bibliography{IEEEabrv,GenSpaceLibTweakingv2.bib}{}
\end{document}